\documentclass{elsart}
\usepackage{psfig}
\usepackage{natbib}

\def\ojet{\Omega_{\rm jet}}

\def\eiso{E_{\rm iso}}
\def\egamma{E_{\gamma}}

\begin{document}
\begin{frontmatter}
\title{A Unified Jet Model of X-Ray Flashes and Gamma-Ray Bursts}
\author[uofc]{D. Q. Lamb}, 
\author[uofc]{T. Q. Donaghy} and 
\author[uofc]{C. Graziani}

\address[uofc]{Department of Astronomy \& Astrophysics, University of Chicago,
				Chicago, IL 60637, USA}

\begin{abstract}
HETE-2 has provided strong evidence that the properties of X-Ray
Flashes (XRFs) and GRBs form a continuum, and therefore that these two
types of bursts are the same phenomenon.  We show that both the
structured jet and the uniform jet models can explain the observed
properties of GRBs reasonably well.  However, if one tries to account
for the properties of both XRFs and GRBs in a unified picture, the
uniform jet model works reasonably well while the structured jet model
does not.  The uniform jet model of XRFs and GRBs implies that most
GRBs have very small jet opening angles ($\sim$ half a degree).  This
suggests that magnetic fields play a crucial role in GRB jets.  The
model also implies that the energy radiated in gamma rays is $\sim$ 100
times smaller than has been thought.  Most importantly, the model
implies that there are $\sim 10^4 -10^5$ more bursts with very small
jet opening angles for every such burst we see.  Thus the rate of GRBs
could be comparable to the rate of Type Ic core collapse supernovae. 
Determination of the spectral parameters and redshifts of many more
XRFs will be required in order to confirm or rule out the uniform jet
model and its implications.  HETE-2 is ideally suited to do this (it
has localized 16 XRFs in $\sim$ 2 years), whereas {\it Swift} is less
so.  The unique insights into the structure of GRBs jets, the rate of
GRBs, and the nature of Type Ic supernovae that XRFs may provide
therefore constitute a compelling scientific case for continuing HETE-2
during the {\it Swift} mission.
\end{abstract}
\end{frontmatter}

\section{Introduction}

Two-thirds of all HETE-2--localized bursts are either ``X-ray-rich'' or
X-Ray Flashes (XRFs); of these, one-third are XRFs \footnote{We define
``X-ray-rich'' GRBs and XRFs as those events for which $\log
[S_X(2-30~{\rm kev})/S_\gamma(30-400~{\rm kev})] > -0.5$ and 0.0,
respectively.} \citep{sakamoto2003b}.  These events have received
increasing attention in the past several years
\citep{heise2000,kippen2002}, but their nature remains largely unknown.

XRFs have $t_{90}$ durations between 10 and 200 sec and their sky
distribution is consistent with isotropy.  In these respects, XRFs are
similar to ``classical'' GRBs.  A joint analysis of WFC/BATSE spectral
data showed that the low-energy and high-energy photon indices of XRFs
are $-1$ and $\sim -2.5$, respectively, which are similar to those of
GRBs, but that the XRFs had spectral peak energies $E_{\rm peak}^{\rm
obs}$ that were much lower than those of GRBs \citep{kippen2002}. The
only difference between XRFs and GRBs therefore appears to be that XRFs
have lower $E_{\rm peak}^{\rm obs}$ values.  It has therefore been
suggested that XRFs might represent an extension of the GRB population
to bursts with low peak energies.

Clarifying the nature of XRFs and X-ray-rich GRBs, and their connection
to GRBs, could provide a breakthrough in our understanding of the
prompt emission of GRBs.  Analyzing 42 X-ray-rich GRBs and XRFs seen by
FREGATE and/or the WXM instruments on HETE-2, \cite{sakamoto2003b}
find that the XRFs, the X-ray-rich GRBs, and GRBs form a continuum in
the [$S_\gamma(2-400~{\rm kev}), E^{\rm obs}_{\rm peak}$]-plane (see
Figure 1, left-hand panel).  This result strongly suggests that these
three kinds of events are the same phenomenon.

Furthermore, \cite{lamb2003c} have placed 9 HETE-2 GRBs with known
redshifts and 2 XRFs with known redshifts or strong redshift
constraints in the ($E_{\rm iso}, E_{\rm peak}$)-plane (see Figure 1,
right-hand panel).  Here $E_{\rm iso}$ is the isotropic-equivalent
burst energy and $E_{\rm peak}$ is the energy of the peak of the burst
spectrum, measured in the source frame.  The HETE-2 bursts confirm the
relation between $E_{\rm iso}$ and $E_{\rm peak}$ found by
\cite{amati2002} for GRBs and extend it down in $E_{\rm iso}$ by a
factor of 300.  The fact that XRF 020903, one of the softest events
localized by HETE-2 to date, and XRF 030723, the most recent XRF
localized by HETE-2, lie squarely on this relation
\citep{sakamoto2003a,lamb2003c} provides additional evidence that XRFs
and GRBs are the same phenomenon.  However, more redshift
determinations for XRFs with 1 keV $< E_{\rm peak} < 30$ keV energy are
needed in order to confirm these results.

\begin{figure}[t]
\centerline{\hbox{
\psfig{file=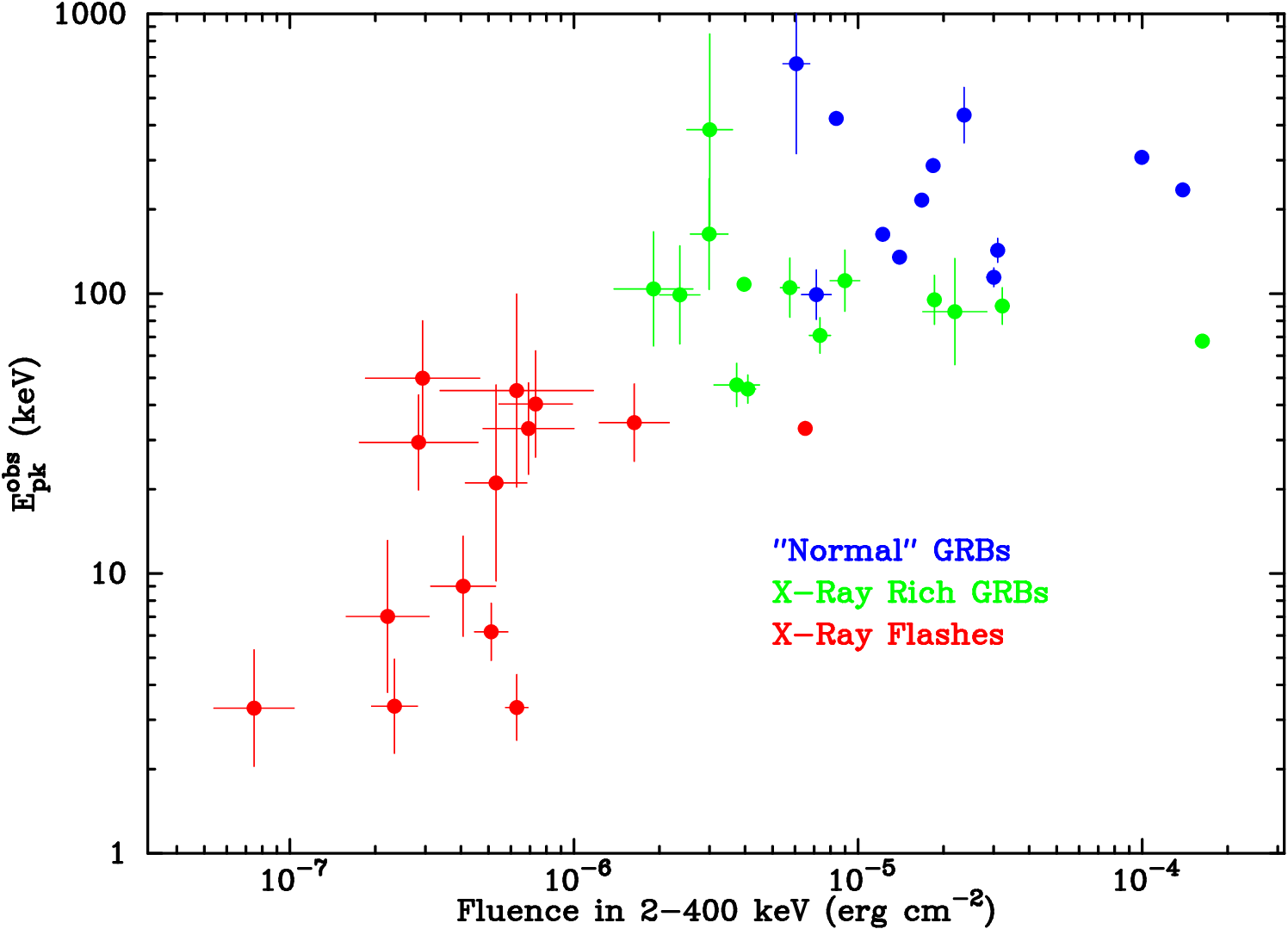,angle=270,width=0.5 \textwidth}
\psfig{file=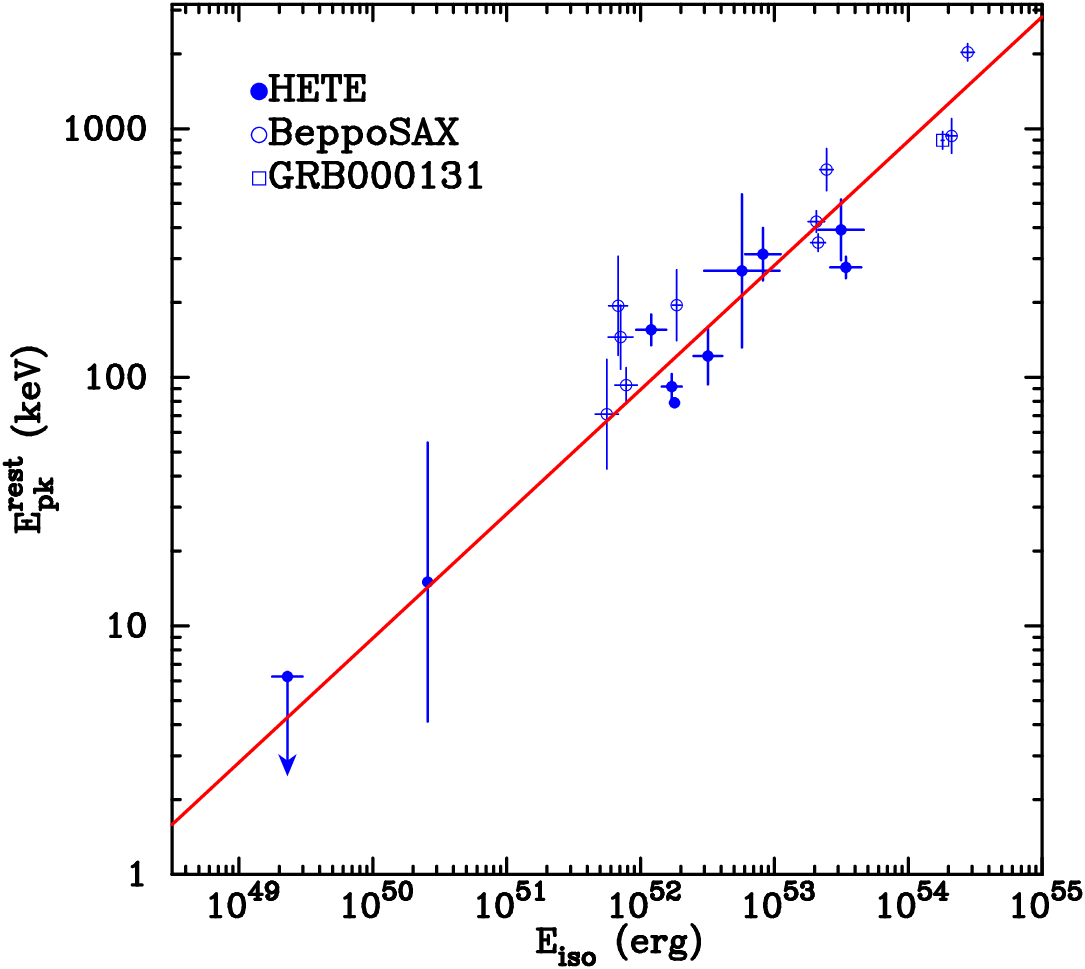,width=0.5 \textwidth}}} 
\caption{Distribution of HETE-2 bursts in the [$S(2-400~{\rm keV}),
E^{\rm obs}_{\rm peak}$]-plane, showing XRFs (red), X-ray-rich GRBs
(green), and GRBs (blue) (left panel).    From \cite{sakamoto2003b}.
Distribution of HETE-2 and BeppoSAX bursts in the ($E_{\rm
iso}$,$E_{\rm peak}$)-plane, where $E_{\rm iso}$ and $E_{\rm peak}$ are
the isotropic-equivalent GRB energy and the peak of the GRB spectrum in
the source frame (right panel).   The HETE-2 bursts confirm the relation
between $E_{\rm iso}$ and $E_{\rm peak}$ found by Amati et al. (2002),
and extend it by a factor $\sim 300$ in $E_{\rm iso}$.  The bursts with
the lowest and second-lowest values of $E_{\rm iso}$ are XRFs 020903
and 030723. From \cite{lamb2003c}.
\label{fig16}}
\end{figure}

Figure 2 shows a simulation of the expected distribution of bursts in
the ($E_{\rm iso}, E_{\rm peak}$)-plane (left panel) and in the 
($F^{\rm peak}_N,E_{\rm peak}$)-plane (right panel), assuming that the
\citep{amati2002} relation holds for XRFs as well as for GRBs
\citep{lamb2003}, as is strongly suggested by the HETE-2 results.  The
SXC, WXM, and FREGATE instruments on HETE-2 have thresholds of $1-6$
keV and considerable effective areas in the  X-ray energy range.  Thus
HETE-2 is ideally suited for detecting and studying XRFs.  In
contrast, BAT on {\it Swift} has a nominal threshold of 20 keV.  This
simulation suggests that the WXM and SXC instruments on HETE-2 detect
many times more bursts with $ E_{\rm peak} < 10$ keV than will BAT on
{\it Swift}.

\section{XRFs as a Probe of GRB Jet Structure, GRB Rate, and Core
Collapse Supernovae}

\cite{frail2001} and \cite{panaitescu2001} [see also \cite{bloom2003}]
have shown that most GRBs have a ``standard'' energy; i.e, if their
isotropic equivalent energy is corrected for the jet opening angle
inferred from the jet break time, most GRBs have the same radiated
energy, $\egamma = 1.3 \times 10^{51}$ ergs, to within a factor of
2-3.

Two models of GRB jets have received widespread attention:

\begin{itemize}

\item
The ``structured jet'' model (see the left-hand panel of Figure 3). In
this model, all GRBs produce jets with the same structure
\citep{rossi2002,woosley2003,zhang2002,meszaros2002}.  The
isotropic-equivalent energy and luminosity is assumed to decrease as
the viewing angle $\theta_v$ as measured from the jet axis increases. 
The wide range in values of $E_{\rm iso}$ is attributed  to differences
in the viewing angle $\theta_v$.  In order to recover the ``standard
energy'' result \citep{frail2001}, $E_{\rm iso} (\theta_v) \sim
\theta_v^{-2}$ is required \citep{zhang2002}.
\bigskip

\item
The ``uniform jet'' model (see the right-hand panel of Figure 3). In
this model GRBs produce jets with very different jet opening angles 
$\theta_{\rm jet}$.  For $\theta < \theta_{\rm jet}$, $E_{\rm iso}
(\theta_v)$ = constant while for $\theta > \theta_{\rm jet}$, $E_{\rm
iso} (\theta_v) = 0$.

\end{itemize}

\begin{figure}[t]
\centerline{\hbox{
\psfig{file=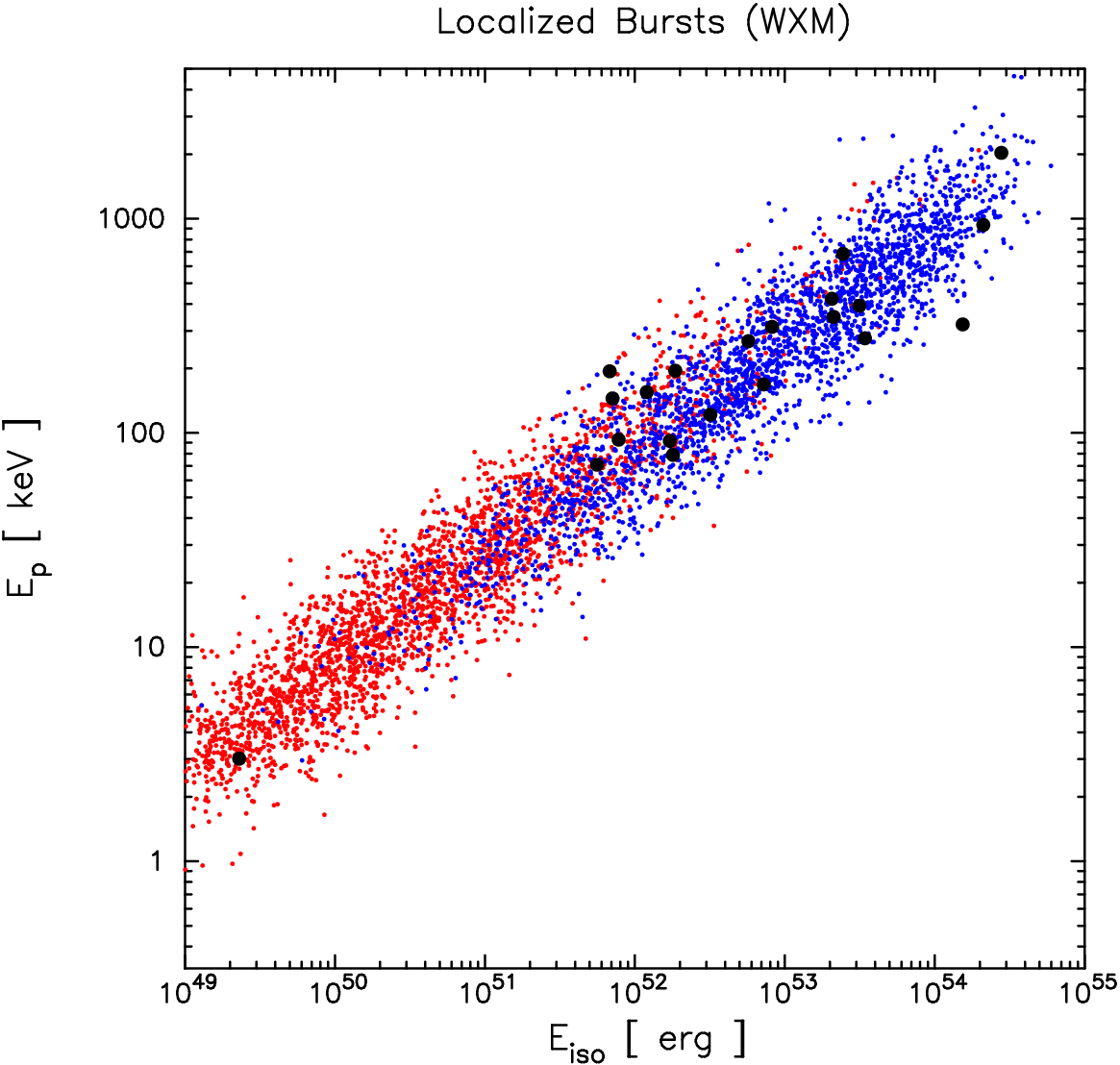,angle=270,width=0.5 \textwidth}
\psfig{file=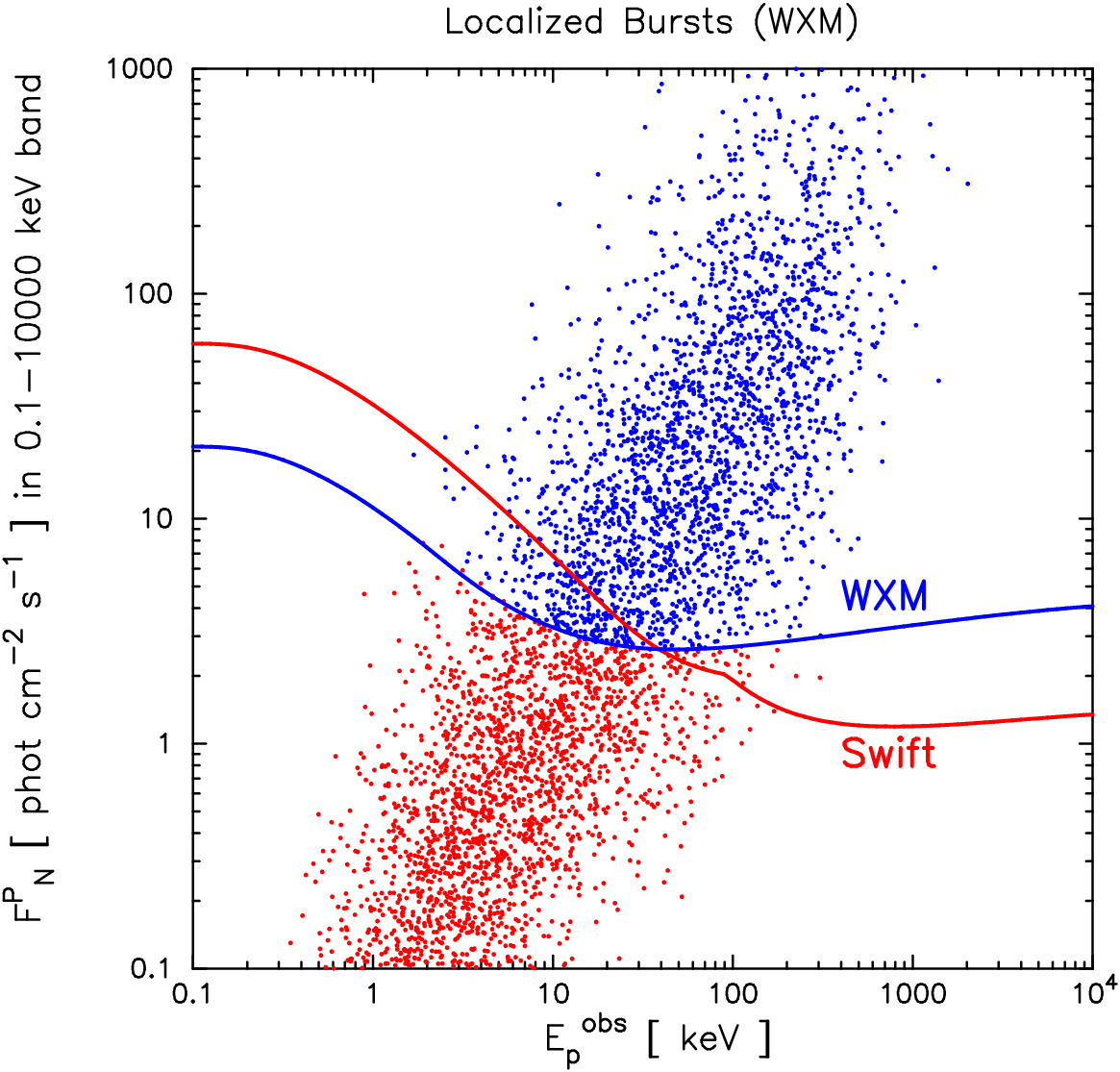,angle=270,width=0.5 \textwidth}}}
\caption{Expected distribution of bursts in the ($E_{\rm iso}, E_{\rm
peak}$)-plane (left panel) and in the  ($F^{\rm peak}_N,E_{\rm
peak}$)-plane (right panel), assuming that the Amati et al. (2002)
relation holds for XRFs as well as for GRBs, as strongly suggested by
the HETE-2 results.  Blue dots are simulated bursts that the WXM on
HETE-2 detects; red dots are simulated bursts that it does not detect. 
The solid dots in the left-hand panel show the locations of HETE-2 and
{\it Beppo}SAX GRBs with known redshifts (the dot at the lower left is
XRF 020903).  The curved lines in the right-hand panel show the
threshold sensitivities of the WXM on HETE-2 and BAT on Swift.  From
\cite{lamb2003}.
\label{fig18}}
\end{figure}

As we have seen, HETE-2 has provided strong evidence that the
properties of XRFs, X-ray-rich GRBs, and GRBs form a continuum, and
that these bursts are therefore the same phenomenon.  If this is true,
it immediately implies that the $\egamma$ inferred by 
\citep{frail2001} is too large by a factor of at least 100
\citep{lamb2003}.  The reason is that the values of $E_{\rm iso}$ for
XRF 020903 \citep{sakamoto2003a} and XRF 030723 \citep{lamb2003c} are
$\sim$ 100 times smaller than the value of  $\egamma$ inferred by Frail
et al. -- an impossibility.

HETE-2 has also provided strong evidence that, in going from XRFs to 
GRBs, $E_{\rm iso}$ changes by a factor $\sim 10^5$ (see Figure 1,
right-hand panel).  If one tries to explain only the range in $E_{\rm
iso}$ corresponding to GRBs, both the uniform jet model and the
structured jet model work reasonably well.  However, if one tries to
explain the range in $E_{\rm iso}$ of a factor $\sim 10^5$ that is
required in order to accommodate both XRFs and GRBs in a unified
description, the uniform jet works reasonably well while the structured
jet model does not.

\begin{figure}[t]
\centerline{\psfig{file=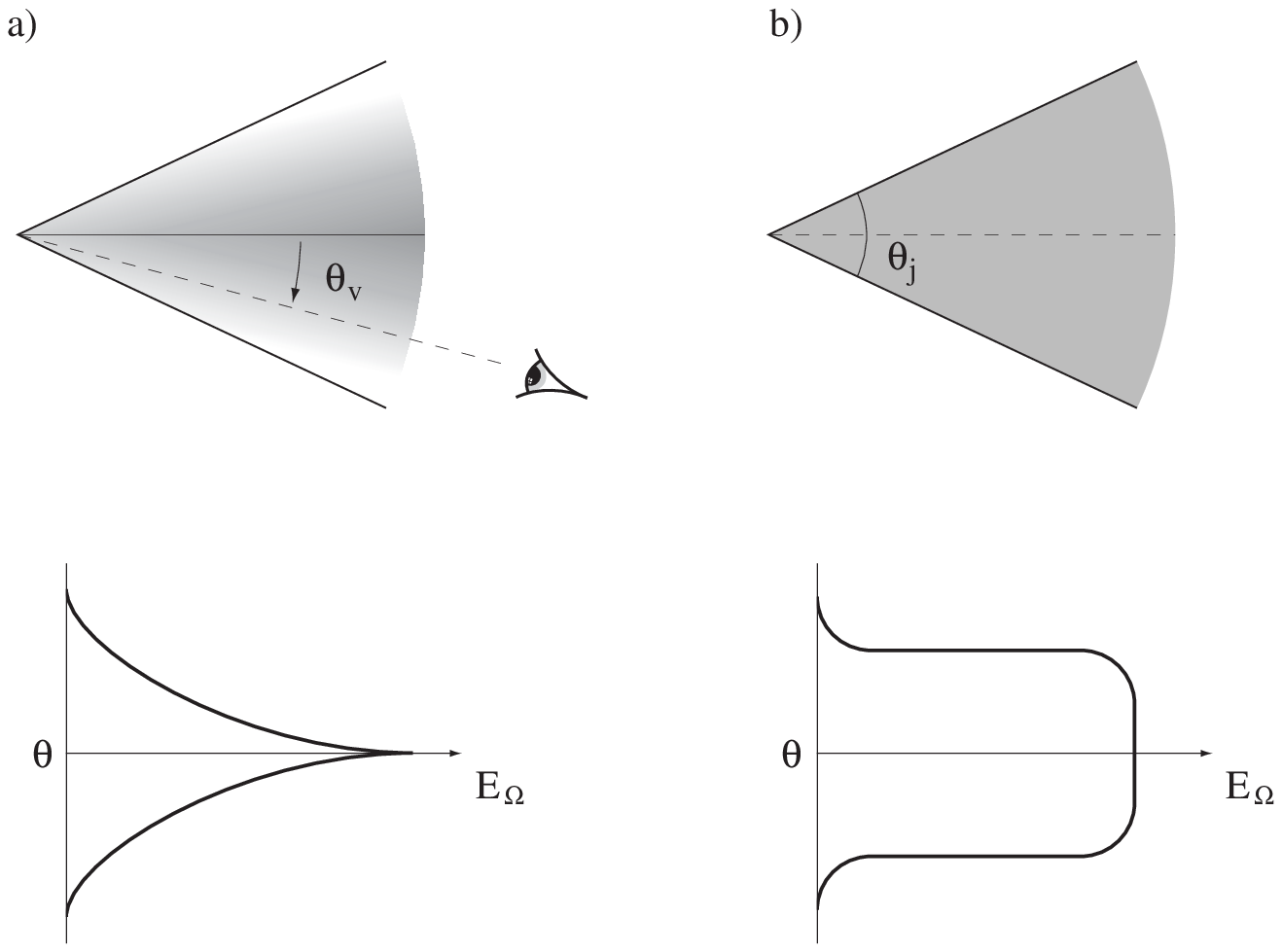,width=0.7 \textwidth}} 
\caption{
Schematic diagrams of the universal and uniform jet models of GRBs
\citep{ramirez-ruiz2002}. 
In the universal jet model, the isotropic-equivalent energy and
luminosity is assumed to decrease as the viewing angle $\theta_v$ as
measured from the jet axis increases.  In order to recover the
``standard energy'' result \citep{frail2001}, $E_{\rm iso} (\theta_v) \sim
\theta_v^{-2}$ is required.  In the uniform jet model, GRBs produce
jets with a large range of jet opening angles  $\theta_{\rm jet}$.  For
$\theta < \theta_{\rm jet}$, $E_{\rm iso} (\theta_v)$ = constant while
for $\theta > \theta_{\rm jet}$, $E_{\rm iso} (\theta_v) = 0$.  
\label{fig20}}
\vskip -0.1truein
\end{figure}

The reason is the following:  the observational implications of the
structured jet model and the uniform jet model differ dramatically if
they are required to explain XRFs and GRBs in a unified picture.  In
the structured jet model, most viewing angles $\theta_v$ are $\approx
90^\circ$.  This implies that the number of XRFs should exceed the
number of GRBs by many orders of magnitude, something that HETE-2 does
not observe (see Figures 1, 2, 4, and 5).  On the other hand, by
choosing $N(\Omega_{\rm jet}) \sim \Omega_{\rm jet}^{-2}$, the uniform
jet model predicts equal numbers of bursts per logarithmic decade in
$E_{\rm iso}$ (and $S_E$), which is exactly what HETE-2 sees (again,
see Figures 1, 2, 4, and 5) \citep{lamb2003}.

Thus, if $E_{\rm iso}$ spans a range $\sim 10^5$, as the HETE-2 results
strongly suggest, the uniform jet model can provide a unified picture
of both XRFs and GRBs, whereas the structured jet model cannot.  XRFs 
may therefore provide a powerful probe of GRB jet structure.

\begin{figure}[t]
\centerline{\hbox{
\psfig{file=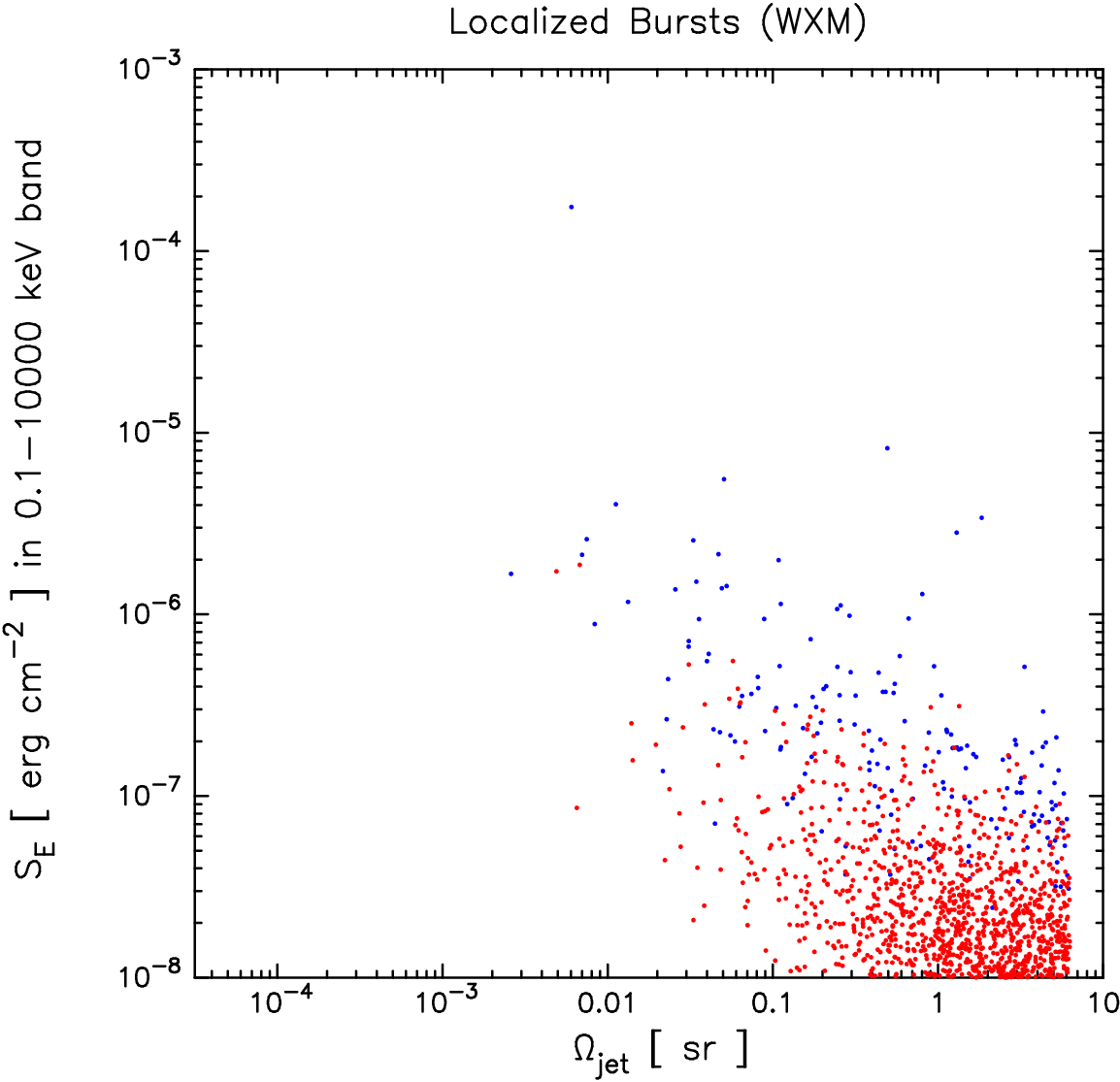,angle=270,width=0.5 \textwidth}
\psfig{file=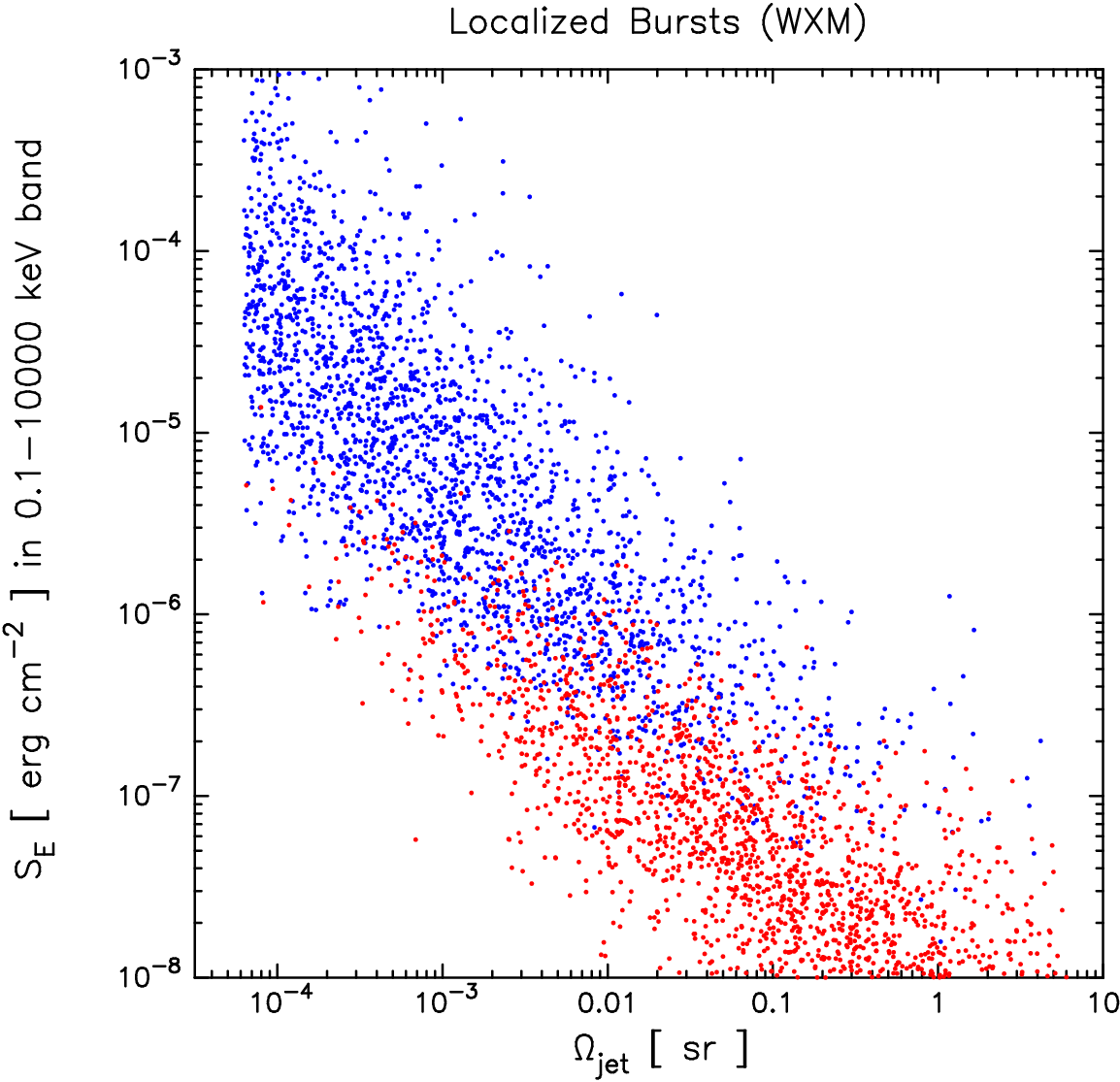,angle=270,width=0.5 \textwidth}}}
\caption{Expected distribution of bursts in the ($\Omega_{\rm
jet},S_E$)-plane for the universal jet model (left panel) and uniform
jet model (right panel), assuming that the Amati et al. (2002) relation
holds for XRFs as well as for GRBs, as the HETE-2 results strongly
suggest.  From \cite{lamb2003}.
\label{fig21}}
\end{figure}

A range in $\eiso$ of $10^5$ requires a {\it minimum} range in $\Delta
\ojet$ of $10^4 - 10^5$ in the uniform jet model.  Thus the unified
picture of XRFs and GRBs based on the uniform jet model implies that
there are $\sim 10^4 - 10^5$ more bursts with very small $\ojet$'s for
every such burst we see; i.e., the rate of GRBs may be $\sim 100$ times
greater than has been thought.  

Since the observed ratio of the rate of Type Ic SNe to the rate of
GRBs in the observable universe is $R_{\rm Type\ Ic}/ R_{\rm GRB} \sim
10^5$ \citep{lamb1999}, a unified picture of XRFs and GRBs based on the
uniform jet model implies that the rate of GRBs could be comparable to
the rate of Type Ic SNe \citep{lamb2003}.  More spherically symmetric
jets yield XRFs and narrow jets produce GRBs.  Thus XRFs and GRBs may
provide a combination of GRB/SN samples that would enable astronomers
to study the relationship between the degree of jet-like behavior of
the GRB and the properties of the supernova (brightness, polarization
$\Leftrightarrow$ asphericity of the explosion, velocity of the
explosion $\Leftrightarrow$ kinetic energy of the explosion, etc.). 
GRBs may therefore provide a unique laboratory for understanding Type
Ic core collapse supernovae.

\begin{figure}[t]
\centerline{\hbox{
\psfig{file=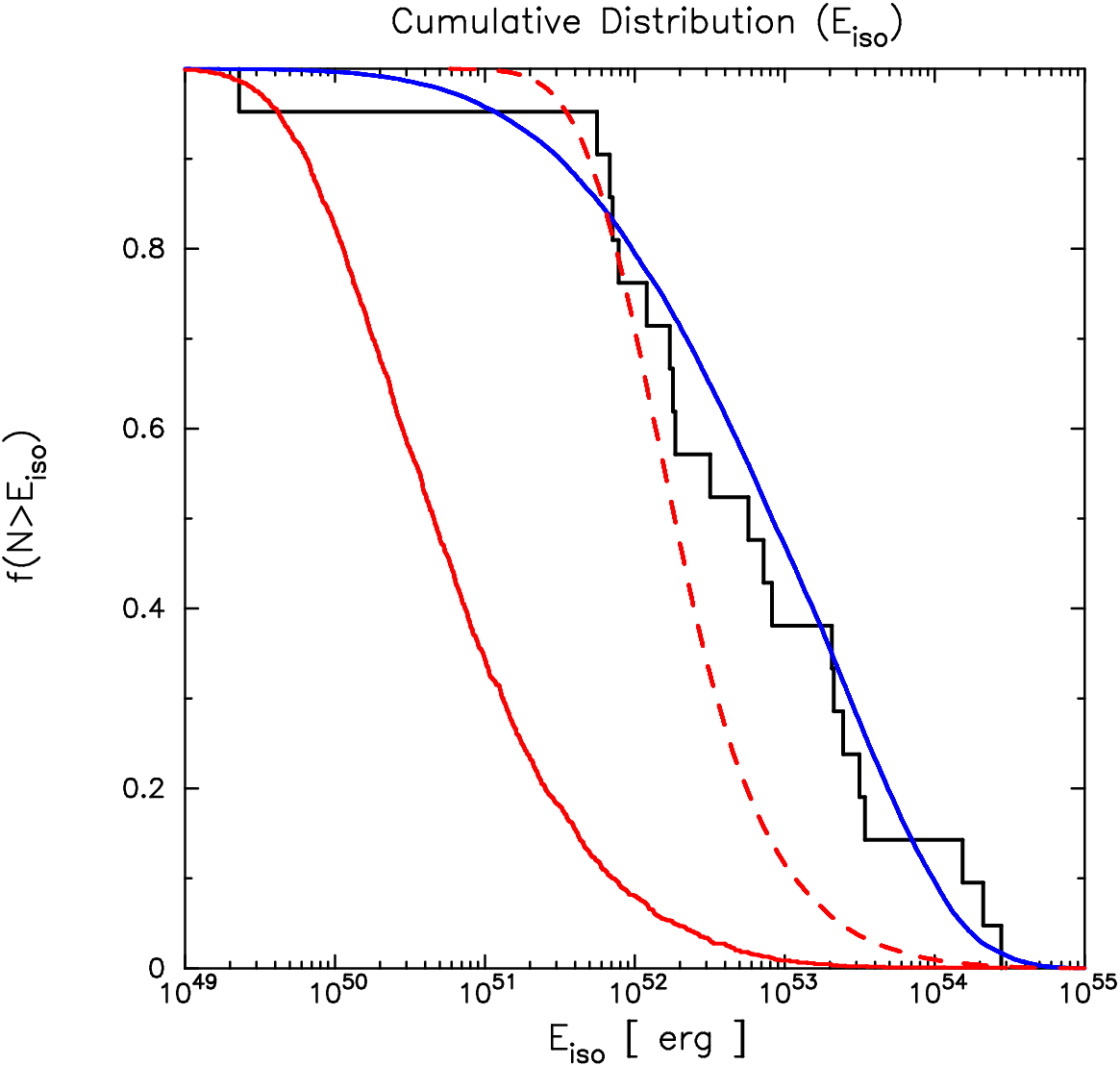,angle=270,width=0.5 \textwidth}
\psfig{file=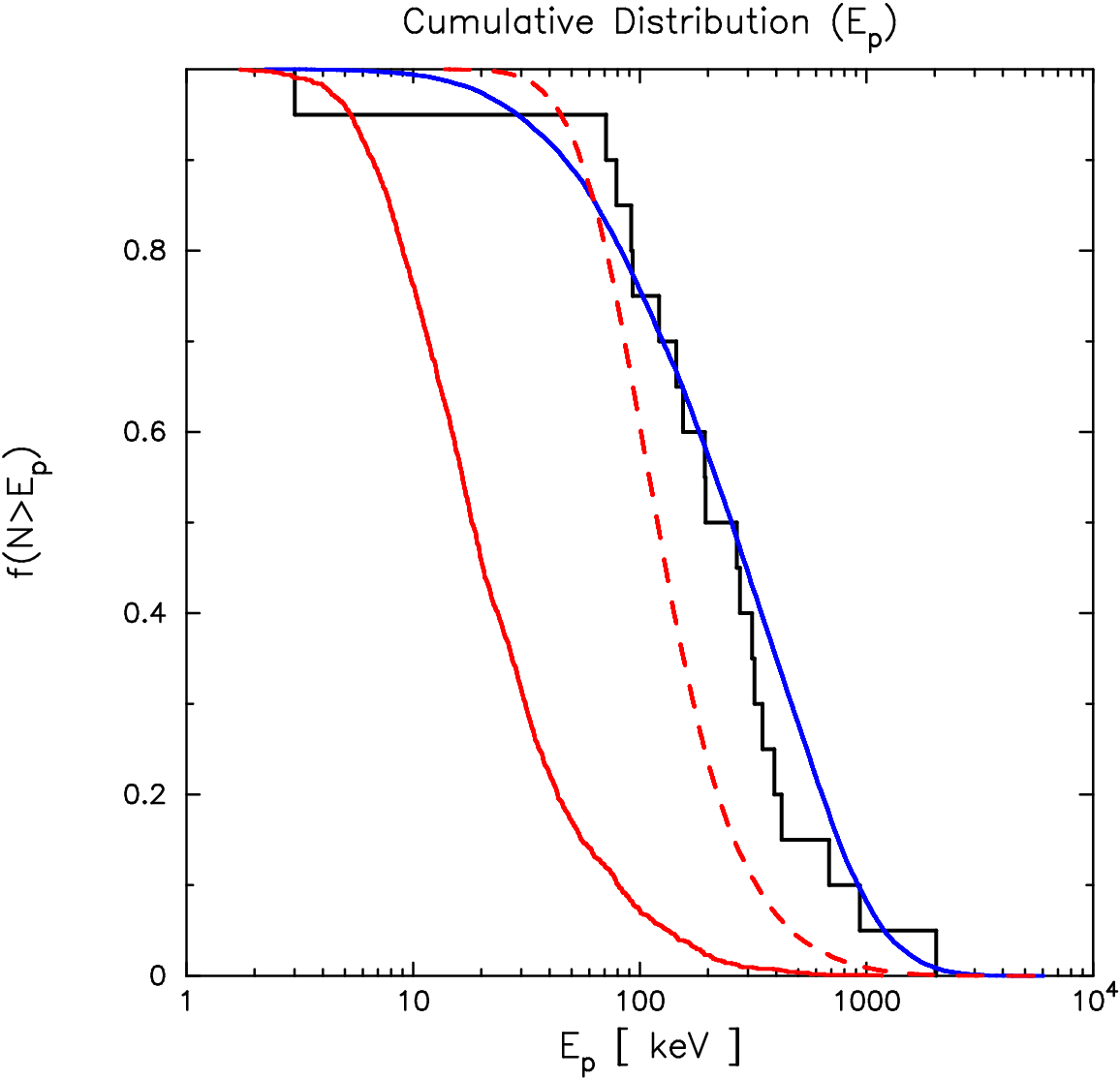,angle=270,width=0.5 \textwidth}
}}
\centerline{\hbox{
\psfig{file=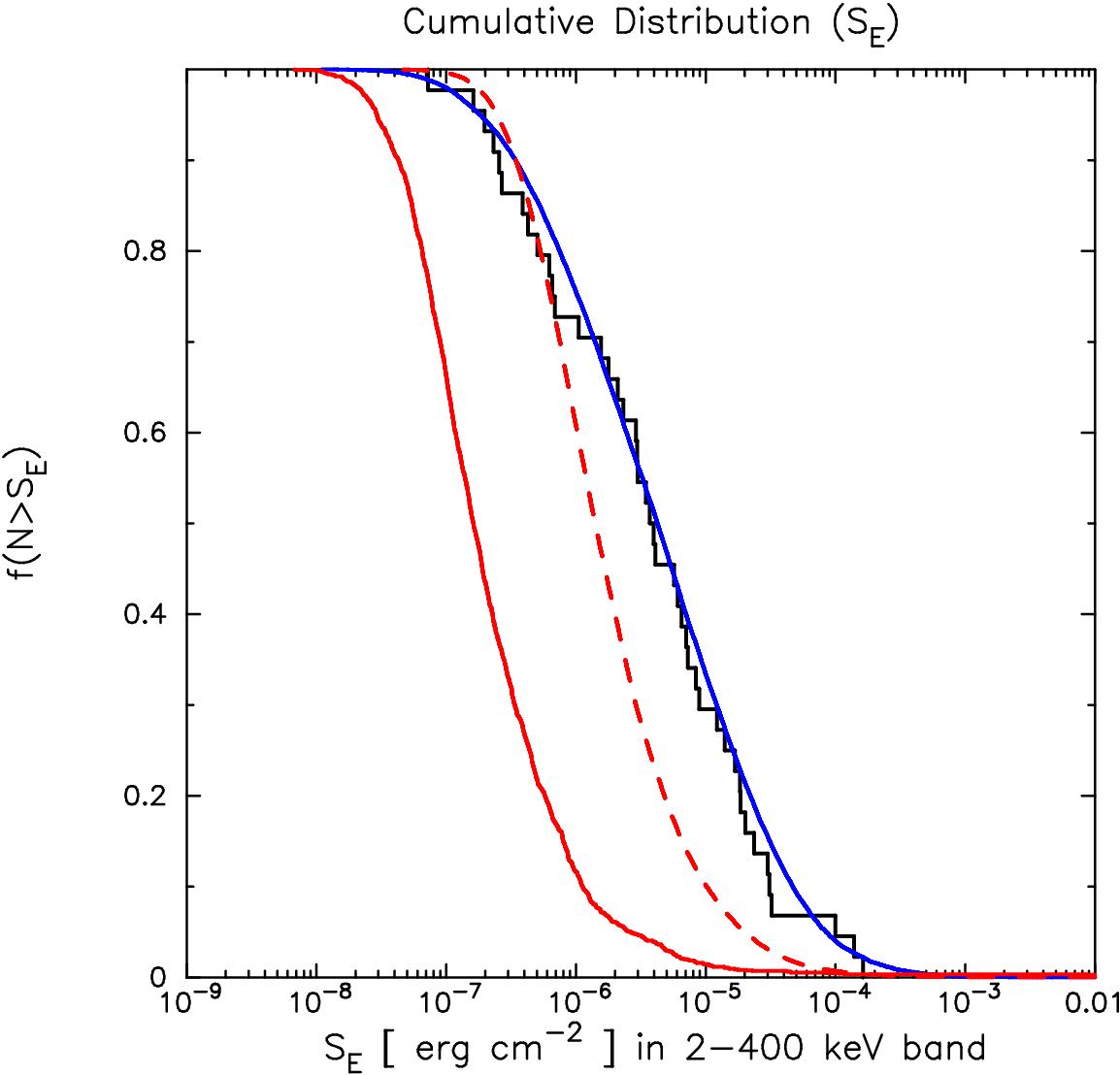,angle=270,width=0.5 \textwidth}
\psfig{file=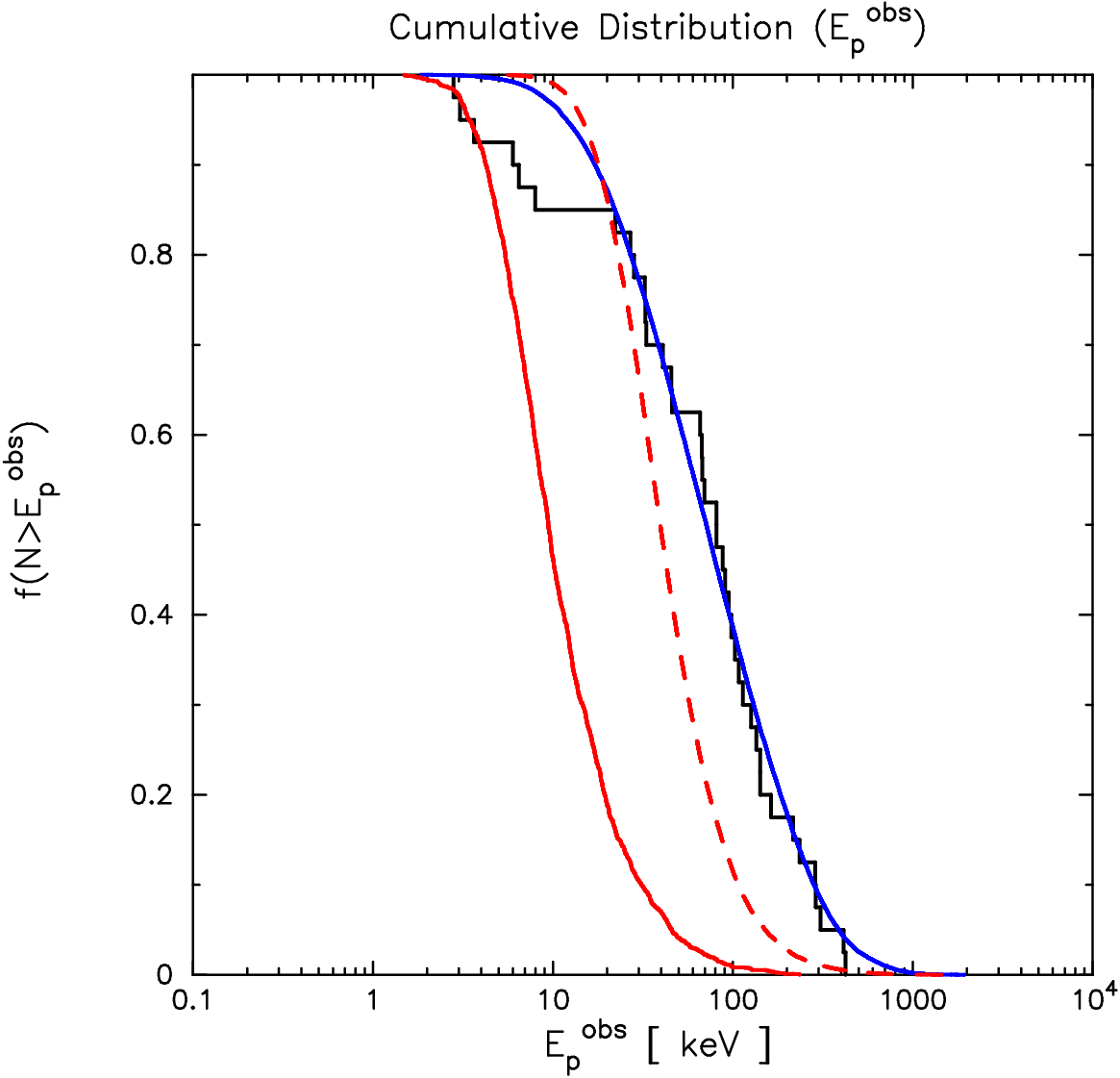,angle=270,width=0.5 \textwidth}
}}
\caption{Top row: cumulative distributions of $S(2-400 {\rm keV})$ (left panel)
and $E^{\rm obs}_{\rm peak}$ (right panel) predicted by the structured
(red) and uniform (blue) jet models, compared to the observed
cumulative distributions of these quantities.  Bottom row: cumulative
distributions of $E_{\rm iso}$ (left panel) and $E_{\rm peak}$ (right
panel) predicted by the structured (red) and uniform (blue) jet models,
compared to the observed cumulative distributions of these quantities. 
The cumulative distributions corresponding to the best-fit structured
jet model that explains XRFs and GRBs are shown as solid lines; the
cumulative distributions corresponding to the best-fit structured jet
model that explains GRBs alone are shown as dashed lines.  The
structured jet model provides a reasonable fit to GRBs alone but cannot
provide a unified picture of both XRFs and GRBs, whereas the uniform
jet model can.  From \cite{lamb2003}.
\label{fig22}}
\end{figure}

\section{Conclusions}

We have shown that a unified picture of XRFs and GRBs based on the
uniform jet model has profound implications for the structure of GRB
jets, the rate of GRBs, and the nature of Type Ic supernovae. 
Obtaining the evidence needed to confirm or rule out the uniform jet
model and its implications will require the determination of both the
spectral parameters and the redshifts of many more XRFs.  The broad
energy range of HETE-2 (2-400 keV) means that it is able to accurately
determine the spectral parameters of the XRFs that it detects and
localizes.  This will be more difficult for {\it Swift}, which has a
more limited spectral coverage (15-140 keV).
Until very recently, only one XRF (XRF 020903; Soderberg et al. 2002)
had a probable optical afterglow and redshift.  This is because the
X-ray (and therefore the optical) afterglows of XRFs are $\sim 10^3$
times fainter than those of GRBs \citep{lamb2003}.  But this challenge
can be met: the recent HETE-2--localization of XRF 030723 represents
the first time that an XRF has been localized in real time
\citep{prigozhin2003}; identification of its X-ray and optical
afterglows rapidly followed \citep{fox2003c}.  This suggests that {\it
Swift}'s ability to rapidly follow up GRBs with the XRT and UVOT --
its revolutionary feature -- will greatly increase the fraction of
bursts with known redshifts.  

Therefore a partnership between HETE-2 and {\it Swift}, in which HETE-2
provides the spectral parameters for XRFs, and {\it Swift} slews to the
HETE-2--localized XRFs and provides the redshifts, can provide the data
that is needed to confirm or rule out the uniform jet model and its
implications.  This constitutes a compelling scientific case for
continuing HETE-2 during the {\it Swift} mission.


\begin{thebibliography}{999}

\bibitem[Amati et al.(2002)]{amati2002} 
        Amati, L., et al. 2002, A \& A, 390, 81
	\bibitem[Band(2003)]{band2003} 
        Band, D. L. 2003, ApJ, in press  (astro-ph/0212452)
\bibitem[Bloom, Frail \& Kulkarni(2003)]{bloom2003} 
        Bloom, J., Frail, D. A. \& Kulkarni, S. R. 2003, ApJ, 588, 945
\bibitem[Fox et al.(2003c)]{fox2003c}
	Fox, D. W., et al. 2003c, GCN Circular 2323
\bibitem[Frail et al.(2001)]{frail2001}
	Frail, D. et al. 2001, ApJ, 562, L55
\bibitem[Heise et al.(2000)]{heise2000}  
        Heise, J., in't Zand, J., Kippen, R. M., \& Woods, P. M., 
	2000, in Proc. 2nd Rome Workshop:  Gamma-Ray Bursts in the
	Afterglow Era, eds. E. Costa, F. Frontera, J. Hjorth (Berlin:
	Springer-Verlag), 16
\bibitem[Kippen et al.(2002)]{kippen2002}
  	Kippen, R. M., Woods, P. M., Heise, J., in't Zand, J., Briggs,
  	M.S., \& Preece, R. D. 2002, in Gamma-Ray Burst and Afterglow
   	Astronomy, AIP Conf. Proc. 662, ed. G. R. Ricker \& R. K.
   	Vanderspek (New York: AIP), 244
\bibitem[Lamb(1999)]{lamb1999} 
	Lamb, D. Q. 1999, A\&A, 138, 607 
\bibitem[Lamb, Donaghy \& Graziani(2003)]{lamb2003} 
        Lamb, D. Q., Donaghy, T. Q., \& Graziani, C. 2003, ApJ, to be
	submitted
\bibitem[Lamb et al.(2003c)]{lamb2003c} 
        Lamb, D. Q., et al. 2003c, to be submitted to ApJ
\bibitem[Lazzati, Ramirez-Ruiz \& Rees(2002)]{lazzati2002}  
        Lazzati, D., Ramirez-Ruiz, E. \& Rees, M. J. 2002, ApJ, 572,
        L57
\bibitem[Lloyd-Ronning, Fryer, \& Ramirez-Ruiz(2002)]{lloyd-ronning2002}
        Lloyd-Ronning, N., Fryer, C., \& Ramirez-Ruiz, E. 2002, ApJ,
	574, 554
\bibitem[M\'esz\'aros, Ramirez-Ruiz, Rees, \& Zhang (2002)]{meszaros2002}
	M\'esz\'aros, P., Ramirez-Ruiz, E., Rees, M. J., \& Zhang, B.
	2002, ApJ, 578, 812
\bibitem[Panaitescu \& Kumar(2001)]{panaitescu2001}
	Panaitescu, A., \& Kumar, P. 2001, ApJ, 556, 1002
\bibitem[Prigozhin et al.(2003)]{prigozhin2003}
	Prigozhin, G., et al. 2003, GCN Circular 2313
\bibitem[Ramirez-Ruiz \& Lloyd-Ronning(2002)]{ramirez-ruiz2002}
	Ramirez-Ruiz, E. \& Lloyd-Ronning, N. 2002, New Astronomy, 7,
	197
\bibitem[Rossi, Lazzati, \& Rees (2002)]{rossi2002}
	Rossi, E., Lazzati, D., \& Rees, M. J. 2002, MNRAS, 332, 945 
\bibitem[Sakamoto et al.(2003a)]{sakamoto2003a}
        Sakamoto, T. et al. 2003a, ApJ, submitted
\bibitem[Sakamoto et al.(2003b)]{sakamoto2003b}
        Sakamoto, T. et al. 2003b, ApJ, to be submitted
\bibitem[Soderberg et al.(2002)]{soderberg2002}
	Soderberg, A. M., et al. 2002, GCN Circular 1554
\bibitem[Woosley, Zhang, \& Heger (2003)]{woosley2003}
	Woosley, S. E., Zhang, W. \& Heger, A. 2003, ApJ, in press
\bibitem[Zhang \& M\'esz\'aros (2002)]{zhang2002}
	Zhang, B. \& M\'esz\'aros, P. 2002, ApJ, 571, 876 
	
\end{thebibliography}
\end{document}